\documentclass{mpe_report}

\usepackage{psfig,graphicx,epsfig}
\usepackage{color}
\usepackage{amsmath,amssymb,epic,eepic,array}

\begin{document}

\pagenumbering{arabic}
\setcounter{page}{120}

\renewcommand{\FirstPageOfPaper }{120}\renewcommand{\LastPageOfPaper }{123}

\title{Study on polarization of high-energy photons from the Crab pulsar}
\author{J. Takata\inst{1},  H.-K. Chang\inst{2}, \and K.S. Cheng\inst{3}}  
\institute{ASIAA/National Tsing Hua University - TIARA,
PO Box 23-141, Taipei, Taiwan
\and Department of Physics and Institute of Astronomy,
 National Tsing Hua University, Hsinchu 30013, Taiwan
\and Department of Physics, University of Hong Kong, Pokfuam Road,
Hong Kong, China}

\maketitle

\begin{abstract}
We investigate polarization of  high-energy emissions from the Crab
pulsar in the frame work of the outer gap accelerator. 
 The recent version of the outer gap, which
extends from inside the null charge surface to the light
cylinder, is used for examining 
 the light curve, the spectrum and the polarization
characteristics, simultaneously. The polarization position angle curve and
the polarization degree are calculated to compare with the
Crab optical data. We show that the outer gap model explains the
general features of the observed light curve, the spectrum and the
polarization by taking into account the emissions from  inside of
the null charge surface and from tertiary pairs, which were produced
by the high-energy photons from the secondary pairs. 
For the Crab pulsar, the polarization position angle curve
indicates
that the viewing angle of the observer measured from the rotational
axis is  greater than $90^{\circ}$.

\end{abstract}

\section{Introduction} 
 Young pulsars such as the Crab pulsar  are strong $\gamma$-ray
sources. The EGRET instrument revealed that the light curve with double peaks in a
period and the spectrum extending to above GeV are typical 
features of the high-energy emissions from the $\gamma$-ray pulsars.  
Although these data have constrained  proposed models,
the origin of the $\gamma$-ray emission is not yet conclusive. On
important reason is that  various models have
successfully explained the features of the observed spectra and/or
light curves. For example, the polar cap model
(Daugherty \& Harding 1996), the caustic model (Dyks et al. 2004) and 
the outer  gap model (Cheng et al. 2000,
hereafter CRZ00), all expect the  main features 
of the observed light curve. So, we cannot
discriminate  the three different models using  the  light curve. 
Furthermore, both polar cap and outer gap models have explained the
observed $\gamma$-ray spectrum (Daugherty \& Harding 1996; Romani
1996).

Polarization measurement will play an
important role to discriminate the various models, because it
increases
observed parameters, namely, polarization degree (p.d.) and position
angle (p.a.) swing. So far, only the optical polarization 
data for the Crab pulsar is available (Kanbach et
al. 2005) in high energy bands. For the Crab pulsar, the spectrum is
 continuously extending from optical to $\gamma$-ray bands. 
In addition, their pulse positions are considered so well that 
the optical emission mechanism is related to 
higher energy emission mechanisms.  In the future, the next generation
Compton telescope  will probably be able to measure 
polarization characteristics in MeV bands. These data
will be useful for discriminating  the different models.

In this paper, we examine the optical polarization characteristics of
the Crab pulsar with the light curve and the spectrum in frame works
of the outer gap model. CRZ00  has
calculated the  synchrotron self-inverse Compton scattering
process of the secondary pairs produced outside  the outer gap 
and has explained the Crab spectrum from X-ray to $\gamma$-ray bands.
In CRZ00, however, the outer-wing 
and the off-pulse emissions of the Crab pulsars cannot be
reproduced, because  the traditional outer gap geometry,  
which extends  from the null charge surface of the
Goldreich-Julian charge density to 
the light cylinder, is assumed. Furthermore, 
 the spectrum in the optical band was not 
considered. In this paper, on these grounds, we modify 
the CRZ00 geometrical model into a more realistic model, following 
 recent 2-D electrodynamical studies (Takata et al. 2004; 
Hirotani 2006), and calculate
 the light curve, the spectrum and the polarization 
characteristics of the Crab pulsar.

\section{Calculation method}
  The outline of the outer gap model for the Crab pulsar is as follows. 
The charge particles are accelerated 
by the electric field  parallel to the
magnetic field lines in  so called gap, where the charge density is
different from the Goldreich-Julina charge density. 
The high energy particles accelerated in the gap 
emit the $\gamma$-ray photons (called primary photons) via the curvature
radiation process. For the Crab pulsar, most of the primary photons
escaping from the outer gap will convert into secondary pairs 
outside the gap, where the accelerating electric field vanishes, 
by colliding with synchrotron X-rays emitted by the secondary pairs. 
The secondary pairs emit  optical - MeV photons via the
synchrotron process  and photons above MeV with
the inverse Compton process.  The high-energy photons emitted by the
secondary pairs may convert into tertiary pairs at higher altitude by
colliding with the soft X-rays from the stellar surface.
 The tertiary pairs emit the optical-UV
photons via the synchrotron process.
 This secondary and tertiary photons appear as the observed 
radiations from the Crab pulsar 

\subsection{Outer gap}
\label{oustruc}
We consider  the outer gap geometry that is extending from inside null
charge surface to near the light cylinder (Takata et al 2004; Hirotani 2006).  
Because the Crab pulsar has a thin gap, we describe the accelerating
electric field (Cheng et al. 1986) with 
\begin{flushleft}
\begin{equation}
E_{||}(r)=\frac{\Omega B(r)f^2(r)R_{lc}^2}{c s(r)},
\label{electric}
\end{equation}
\end{flushleft}
where $f(r)$ is the local gap thickness in units of the light radius,
$R_{lc}=c/\Omega$, and $s(r)$ is the curvature radius of the magnetic
field line.  The typical fractional size of the outer gap is given by 
$f(R_{lc}/2)\sim5.5B^{-4/7}P^{26/21}$ ($\sim 0.04$ for the Crab pulsar)
and the local fractional size is estimated by $f(r)\sim
f(R_{lc}/2)(2r/R_{lc})^{1.5}$ (Zhang \& Cheng 1997).

The electric field described by equation~(\ref{electric})  can be
adopted for the acceleration beyond the null charge surface. 
Inside null charge surface, the electric field rapidly 
 decreases  because of the screening effects of the pairs.  
To simulate the accelerating  electric field inside null charge
surface,  we assume the following field, 
\begin{equation}
E_{||}(r)=E_{n}\frac{(r/r_{i})^2-1}{(r_{n}/r_{i})^2-1},~~r_i\leq r\leq
r_n, 
\label{electeq}
\end{equation}
where $E_{n}$ is the strength of the electric field at the null charge
surface and  $r_{n}$ and $r_{i}$ are the radial distances to the 
null charge surface and the inner boundary of the gap,
respectively.  The local Lorentz
factor of the accelerated particles in the outer gap 
is described by  $\Gamma_e(r)=[3s^2(r)E_{||}/2e]^{1/4}$ with assuming
the force balance between the acceleration and the curvature radiation
back reaction.

We assume that the outer gap extends around the whole polar cap. 
 In the calculation, we constrain the boundaries of the
axial distance and radial distance for the emission regions with
$\rho_{\max}=0.9R_{lc}$ and $r=R_{lc}$, respectively.

\subsection{Synchrotron  radiation from the pairs}
We calculate the synchrotron radiation at each radiating point
following CRZ00. The photon spectrum of the synchrotron
radiation is (CRZ00)
\begin{eqnarray}
F_{syn}(E_{\gamma},r)&=&\frac{3^{1/2}e^3B(r)\sin\theta_{p}(r)}{mc^2h
E_{\gamma}} \nonumber \\
&\times&\int\left[\frac{dn_e(r)}{dE_e}\right]F(x)dE_edV_{rad},
\label{emis}
\end{eqnarray}
where $x=E_{\gamma}/E_{syn}$, $E_{syn}$ is the typical photon energy,
$\theta_p$ is the pitch angle of the particle,
$F(x)=x\int_x^{\infty} K_{5/3}(y)dy$, where $K_{5/3}$ is the modified
Bessel function of order 5/3, $dn_e/dE_{e}\propto E_e^{-2}$ 
is the distribution of the pairs and $dV_{rad}$ is the volume element of
the radiation region considered. 

The pitch angle of the secondary pairs is estimated from 
$\sin\theta_p(R_{lc})\sim \lambda/s(R_{lc})$, where $\lambda$ is the
mean free path of the pair-creation between the primary $\gamma$-rays
and the non-thermal X-rays from the secondary pairs. For the Crab
pulsar,  the mean free path becomes  $\lambda\sim
(n_X\sigma_{\gamma\gamma})^{-1}\sim 10^7\mathrm{cm}$, where we used
the typical number density $n_{X}\sim L_{X}(<E_X>)/\delta\Omega
R_{lc}^2 c <E_X>\sim 8\times10^{17}~\mathrm{cm}^{-3}$, the
luminosity $L_X\sim
10^{35}\mathrm{erg/s}$, the typical soft-photon energy for the
pair-creation, $<E_X>\sim (2m_ec^2)^2/10~\mathrm{GeV}\sim100$~eV, 
and $\sigma_{\gamma\gamma}\sim\sigma_{T}/3$. Therefore the pitch
 angle of the secondary pairs at the
light cylinder is estimated as $\sin\theta_{p}\sim\lambda/s(R_{lc})\sim
0.06$, and the  local pitch angle is calculated from
$\sin\theta_p(r)=\sin\theta_p(R_{lc})(r/R_{lc})^{1/2}$.

Some high-energy photons emitted by the inverse Compton process of the
secondary pairs may convert into tertiary pairs at
higher altitude by colliding  with thermal X-ray photons from the
star. In this paper, we assume that the maximum energy of and  the local
number density of the tertiary pairs are smaller than about $10\%$ of
 those of the secondary pairs. Because the pitch  angle of the
pairs  increases with altitude, we use $\sin\theta_p=0.1$ for the
pitch angle of the tertiary pairs. In fact, the results
 are not sensitive to the pitch angle of the tertiary pairs.

\subsection{Stokes parameters}
For a high Lorentz factor, we can anticipate
 that the emission direction of the particles  coincides with the
direction of the particle's velocity. In  the inertial  
observer frame, the particle motion may be described by 
\begin{equation}
\mbox{\boldmath$n$}=\beta_0\cos\theta_p\mbox{\boldmath$b$}+\beta_0\sin\theta_p\mbox{\boldmath$b$}_
{\perp}+\beta_{co}\mbox{\boldmath$e$}_{\phi},
\label{pmotion}
\end{equation}
where the first term in the right hand side represents the particle
motion along the field line, for which we use the rotating dipole
field,  and  $\mbox{\boldmath$b$}$ 
is the unit vector of the magnetic field line.
The second term in equation (\ref{pmotion}) represents gyration
motion  around the magnetic field line and $\mbox{\boldmath$b$}_{\perp}\equiv\cos\delta\phi\mbox{\boldmath$k$}+\sin\delta\phi\mbox{\boldmath$k$}
\times\mbox{\boldmath$b$}$ is the unit vector perpendicular to the
magnetic field line, where  $\delta\phi$ refers the phase of gyration
motion and $\mbox{\boldmath$k$}=(\mbox{\boldmath$b$}\cdot\nabla)\mbox{\boldmath$b$}/
|(\mbox{\boldmath$b$}\cdot\nabla)\mbox{\boldmath$b$}|$ is the unit
vector of the curvature of the magnetic field lines.  
The third term is co-rotation motion with the star, 
$\beta_{co}=\rho\Omega/c$. The emission direction
of equation (\ref{pmotion})  is described in terms of the
viewing angle measured from the rotational axis, $\xi=\cos^{-1}n_z$, 
and the rotation phase, 
$\Phi=-\Phi_{n}-\mbox{\boldmath$r$}\cdot\mbox{\boldmath$n$}$, where 
$n_z$ is the component of the emission direction parallel to the
rotational axis, $\Phi_{n}$ is the azimuthal angle of the emission
direction  and
$\mbox{\boldmath$r$}$ is the emitting location in units of the light radius.

Because the particles distribute on the gyration phase
$\delta\phi$, the emitted beam
at each point must  become cone like shape with opening angle
$\theta_p (r)$. We calculate the radiations from the particles for all of the 
 gyration phase $\delta\phi= 2\pi i/n$ ($i=1,\cdots,n-1$).

We assume that the radiation at each point  linearly polarizes
with degree of $\Pi_{syn}=(p+1)/(p+7/3)$, where $p$ is the power law
index of the particle distribution, and circular polarization is zero,
that is, $V=0$ in terms of the Stokes parameters. The direction of the
electric vector of the electro-magnetic wave toward the observer 
is parallel to the projected direction of the acceleration 
of the particle on the sky, that is, 
$\mbox{\boldmath$E$}_{em}\propto \mbox{\boldmath$a$}
-(\mbox{\boldmath$n$}\cdot\mbox{\boldmath$a$})\mbox{\boldmath$n$}$ 
(Blaskiewicz et al. 1991). In the present
case, the acceleration with  equation (\ref{pmotion}) 
is approximately written by $\mbox{\boldmath$a$}\sim
\beta_{0}\omega_B\sin\theta_p(-\sin\delta\phi\mbox{\boldmath$k$}
+\cos\delta\phi\mbox{\boldmath$k$}\times\mbox{\boldmath$b$})$.

 We define the position angle $\chi^i$ to be angle between 
the electric field $\mbox{\boldmath$E$}_{em}$  and the projected
rotational axis on the sky.
The Stokes parameters $Q^{i}$ and
$U^{i}$ at each radiation is represented by
$Q^{i}=\Pi_{syn}I^{i}\cos2\chi^{i}$ and
$U^{i}=\Pi_{syn}I^{i}\sin2\chi^{i}$. After collecting the photons from
the possible points for
each rotation phase $\Phi$ and  a viewing angle $\xi$, the expected
p.d. and p.a. are, respectively,  obtained from 
$P(\xi,\Phi)=\Pi_{syn}\sqrt{Q^2(\xi,\Phi)
+U^2(\xi,\Phi)}/I(\xi,\Phi)$ and $
\chi(\xi,\Phi)=0.5\mathrm{atan}
\left[U(\xi,\Phi)/Q(\xi,\Phi)\right]$. 

The  inclination angle $\alpha$
 and the viewing angles $\xi$ measured from the rotational axis 
are the  model  parameters. 
The  radial distance $r_i$ of the inner boundary in equation
(\ref{electeq}) is also a model  parameter, because  
the distance $r_i$ is determined
by the current through the gap (Takata et al. 2004).  Because the
last-open line must be modified from the traditional magnetic surface,
which is tangent to the light cylinder for the vacuum case, 
by the plasma effects (Romani
1996), the altitude of the upper surface of the outer gap, above which
the pairs are created and emit the synchrotron photons, is used as a
model parameter using a fractional polar angle $a=\theta_u/\theta_{lc}$,
 where $\theta_u$ is the polar angle of the footpoints of the
magnetic field lines of the gap upper surface and $\theta_{lc}$ is 
the polar angle of the field lines which are tangent to the light
cylinder for the vacuum case.  

\section{Results}
\begin{figure}
\centerline{\psfig{file=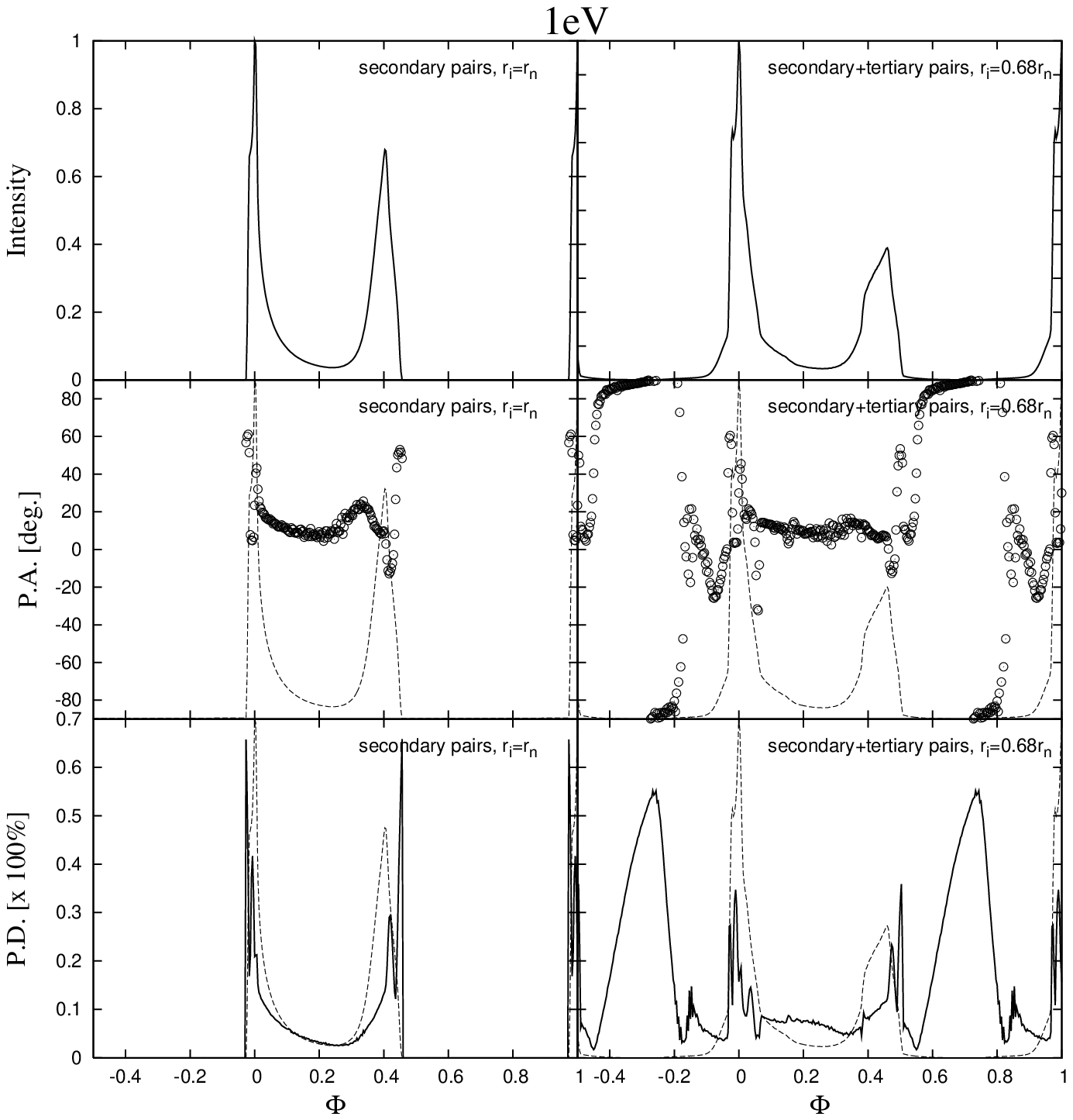,width=7.5cm,height=4.0cm,clip=} }
\caption{Polarization characteristics for
the traditional model (left column) and the present model (right
column).
 The upper, middle and lower
panels in each column show, respectively, the light curve, the
position angle and the polarization degree for
$\alpha=40^{\circ}$ and $a=0.93$.
\label{compari}}
\end{figure}
Figure~\ref{compari} compares the polarization characteristics at
1~eV predicted by the traditional (left column) 
and the present (right column) outer gap models. The traditional model
considers the emissions from the secondary pairs with the outer gap
starting from the null charge surface of the Goldreich-Julian charge
density, $r_i=r_n$. The present model takes into account the emissions
from inside null charge surface and the tertiary pairs. We assume the
outer gap starts from the radial distance of 68\% of the distance to
null charge surface,  $r_i=0.68r_n $. 
The model parameters are $\alpha=40^{\circ}$, $a=0.93$ and $\xi\sim101^{\circ}$,
 where the viewing angle is
chosen so that the predicted phase separation between the two peaks is
consistent with the observed value $\delta\Phi\sim0.4$~phase. 

From the pulse profiles, we find that the traditional model can not
produce the outer-wing and the off-pulse emissions. On the other hand,
the present model produces the  outer-wing and the off-pulse
emissions with the emissions from  inside of the null charge
surface. 

As  seen in the polarization characteristics by the traditional model,
we find that the secondary emissions beyond the
null charge surface  make the polarization characteristics  such  
that the polarization degree takes a lower value 
at the bridge phase and a larger value near the peaks. In the
synchrotron case, the cone like beam is radiated at each point, and an
overlap of the radiations from the different  particles on the
gyration phase causes the depolarization. For the viewing angle
$\xi\sim101^{\circ}$,  the radiations from all gyration phases
contribute to the observed radiation at the bridge phase, because the
line of sight passes through middle part of the emission regions.  
In such a case, the depolarization is strong, and as a result, the
emerging radiation from the secondary  pairs polarizes with  a very
low p.d. ($<10\%$).  Near the peaks, 
the radiations from the some range of gyration phases are not observed because
the line of sight passes through the edge of the emission regions at
peaks. In such a case, the depolarization is
weaker and the emerging  radiation  highly polarizes.

In the polarization angle swing in the present model, the large swings
 appear at the both peaks, and the difference of the position angle
between the off-pulse and the bridge phases is about 90~degree. 
We can see the effects of the tertiary pairs on the polarization
characteristics at the bridge phase.  
By comparing the p.d. at the bridge phase,  
we find that tertiary pairs produce the radiations with 
 $\sim 10\%$ of the p.d. at the bridge phase. 

\begin{figure}
\centerline{\psfig{file=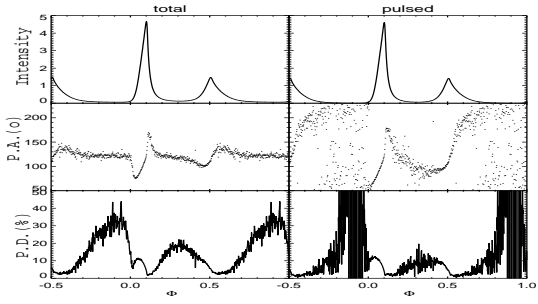,width=7.5cm,height=4.0cm, clip=} }
\caption{Optical polarization for the Crab pulsar. Left:Polarization 
characteristics for the total emissions. Right:Polarization characteristics of the emissions after
subtraction of the DC level (Kanbach et al. 2005).
 The figures was transcribed from  Dyks et al. (2004).
\label{data}}
\end{figure}

Figure~\ref{data} summarizes  the Crab optical data. 
Left column and the right column show, respectively, the Crab optical data for
the total emissions and for the emissions after subtraction of the DC level, 
which has the constant intensity at the
level of 1.24\% of the main pulse intensity. 
In the total emissions (left column), the impressive
polarization feature is that the off-pulse and
bridge phases have the  fixed value of the p.a. These polarization features
of the observation are not predicted by the present model. The present
model are more consistent with the Crab optical data 
after the subtraction of the DC level.  
Especially, the model reproduces the most striking feature in
 the observed p.a. that the large swing at both peaks, and 
the  observed low p.d. at bridge phase $\sim 10\%$. Also, 
the pattern of the p.d. are reproduced by the present model. 

\begin{figure}
\centerline{\psfig{file=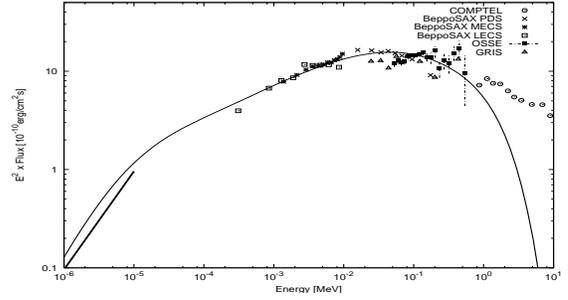,width=7.5cm,height=4.0cm,clip=} }
\caption{The optical-X ray spectrum for the Crab pulsar. The
calculation is for $\alpha=40^{\circ}$, $a=0.93$, $r_i=0.68$ 
and $\xi\sim101^{\circ}$.  The X-ray data  are taken
from Kuiper et al. (2002) and reference therein,  
and the optical data from Sollerman et al. (2000).
\label{spectrum}}
\end{figure}

Figure~\ref{spectrum} compares the model spectrum with the Crab data 
in optical-MeV bands.  The model parameters are same with that in the
right column in Figure~\ref{data}. In this case, we assume 
that the pairs escape from the light
cylinder with the Lorentz factor $\Gamma\sim20$ due to the synchrotron
cooling effect, which  makes a spectral
break around 10~eV in Figure~\ref{spectrum}.
The model spectrum also explains the general features of
the data. The outer gap model can explain the general
features of the observed light curve, the spectrum and the
polarization characteristics in optical band for the Crab pulsar,
simultaneously.

It is worth to note that we can distinguish the two viewing angles
mutually symmetric with respect to the rotational equator using the
polarization swing.  For such symmetric viewing angles,   
the  light curves, the spectra and the p.d. curves 
are identical.  However, the p.a. curves  are mirror symmetry 
with respect to the rotational equator because of the difference directions of 
the projected magnetic field on the sky. The pattern of the position 
angle  for the viewing angle smaller 
than $90^{\circ}$ swings to opposite directions from the Crab data 
at the both peaks. Therefore, the present model predicts that the
viewing angle 
larger than $90^{\circ}$  are preferred for the Crab pulsar.

\section{Conclusions}
We have considered the light curve,  the spectrum and the polarization
characteristics for the Crab
pulsar predicted by the outer gap model which takes into account 
the emissions from inside the null charge surface and from the
tertiary pairs. We have shown that 
the expected polarization characteristics are consistent with
the Crab optical data after subtraction of the DC level. The outer gap
model explains the spectrum, light curve and the polarization
characteristics, simultaneously.  

\begin{acknowledgements}
 This work was supported
by the Theoretical Institute for Advanced Research in Astrophysics
(TIARA) operated under Academia Sinica and the National Science 
Council Excellence Projects program
in Taiwan administered through grant number NSC 94-2112-M-007-002,
NSC 94-2752-M-007-002-PAE and  NSC 95-2752-M-007-001-PAE. And the
author, KSC, was supported by  a RGC grant number HKU7015/05P.

\end{acknowledgements}
   


            \clearpage

\end{document}